\documentclass[11pt]{article}

\usepackage{amsfonts}
\usepackage{amsmath}
\usepackage{amssymb}
\usepackage{amsthm}
\usepackage[total={7in,9in}]{geometry}
\usepackage{graphicx}
\usepackage{lmodern}
\usepackage{hyperref}
\usepackage[shortlabels]{enumitem}

\usepackage[english]{babel}

\title{Valida ISA Spec, version 1.0 \smallskip
  \\ \large A zk-Optimized Instruction Set Architecture}
\author{
     Morgan Thomas\textsuperscript{1}, Mamy Ratsimbazafy\textsuperscript{1},
     Marcin Bugaj\textsuperscript{1}, Lewis Revill\textsuperscript{1},
  \\ Carlo Modica\textsuperscript{1}, Sebastian Schmidt\textsuperscript{1},
     Ventali Tan\textsuperscript{1},
  \\ Daniel Lubarov \textlangle\texttt{daniel.l@polygon.technology}\textrangle,
  \\ Max Gillett\textlangle\texttt{max.gillett@gmail.com}\textrangle, Wei Dai \textlangle\texttt{me@wdai.us}\textrangle
  \\ \\ \textsuperscript{1} Lita, global \\ \texttt{\{morgan,mamy,marcin,lewis,carlo,sebastian,ventali\}@lita.foundation}
}

\begin{document}

\maketitle
\begin{abstract}
The Valida instruction set architecture is designed for implementation in zkVMs to optimize for fast, efficient execution proving. This specification intends to guide implementors of zkVMs and compiler toolchains for Valida. It provides an unambiguous definition of the semantics of Valida programs and may be used as a starting point for formalization efforts.
\end{abstract}

\tableofcontents

\section{Motivation}

Succinct proofs of execution (SPEs) are a way of implementing verifiable computation which balances efficiency with ease of implementation. \cite{zexe} SPEs prove the result of executing a program, without revealing the exact input or requiring the verifier to re-run the computation. Verifiable computation using SPEs is important for ensuring the integrity of blockchain protocols. For example, SPEs are involved in the designs of ZK rollups \cite{starkex,zksync}, but also increasingly in the designs of optimistic rollups, with the rise of ZK fraud proofs \cite{opsuccinct,yu24}.

Succinct proofs of complex facts are challenging to create due to their resource-intensive nature and the complexity of the algorithms for making them. The motivation for SPEs is to simplify the process of creating succinct proofs. Given the ability to create SPEs for programs running on a virtual machine, usually called a zkVM,\footnote{The term ``zkVM'' as commonly applied is often a misnomer in that many zkVM projects do not actually attempt to enforce the zero-knowledge property which is the namesake of zkVM, short for ``zero-knowledge virtual machine.''} it reduces the problem of creating succinct proofs to the problem of creating programs to run on the zkVM. The need to create succinct proofs of execution (SPEs) gives rise to challenges and opportunities in implementing programming language toolchains. 

The challenge with applying zkVM implementations, traditionally, has been that creating the proofs of execution is computationally intensive. The unique nature of succinct proofs gives rise to unique opportunities to optimize the execution of programs in the zkVM environments. The zkVM environment itself can be optimized, but so can the programming language toolchains used to target it, and the instruction set architecture which the zkVM is designed to run.

Since the zkVM execution environment is substantially different from a hardware execution environment, a different cost model applies. This means that optimal code generation strategies and instruction set architectures will be different for a zkVM execution environment as compared to a hardware execution environment. These observations led to various attempts to make ISAs and programming language toolchains which are optimized for making SPEs. These include Cairo \cite{cairo}, Leo \cite{leo}, and Valida \cite{validadocs,validaspec}.

Lita has been developing a Valida zkVM and associated compiler toolchain since 2023. The Valida compiler toolchain supports compiling Rust, C, WASM, and LLVM IR to Valida machine code. The research, testing, and user feedback which Lita has performed and received during that time indicate that Valida is an excellent choice of ISA for fast and efficient succinct execution proving. \cite{thomas25} Lita's research has also indicated that some changes to the Valida ISA as specified in \cite{validaspec} are not only beneficial for performance, but also needed for functional completeness of the execution environment.

This spec reflects all of the changes to the basic Valida ISA made by Lita. The reason for creating and publishing this spec is to support efforts to implement Valida and compiler toolchains for Valida. A precise understanding of Valida's semantics is needed to ensure correctness of zkVMs and compiler toolchains based on Valida. This spec therefore provides such a rigorous definition of Valida's semantics, in a form which is sufficient as a starting point for formalization.

\subsection{Why Valida is efficient for succinct proofs}

Why build a zkVM based on Valida, rather than a more established standard such as RISC-V? \emph{(What follows in this section is re-printed with light editing from \cite{thomas25}.)}

For purposes of efficiently making succinct proofs of execution, Valida is a better starting point than RISC-V. The Valida ISA was designed specifically for making SPEs. The main difference is that RISC-V has a bank of 31 general-purpose registers, whereas Valida has no general-purpose registers and instead, most Valida opcodes directly address stack operands held in RAM. As a result, Valida programs do not need instructions to deal with register spilling or saving and restoring register values at function call boundaries. This allows compilers to generate Valida code which in many cases executes fewer instructions than would be needed in the case of RISC-V. Executing fewer instructions tends to correlate with more efficient proving.

In a CPU, a register is a memory location that is located relatively close to the control unit and arithmetic logic unit (ALU), offering relatively fast read-write latency. A register holds a relatively small amount of data: typically one word, a small number of words, or as little as one bit. A general-purpose register is typically used for holding inputs and outputs of arithmetic and logical operations.

Registers are the lowest-latency form of volatile memory, with the least storage capacity. The next lowest-latency form of volatile memory is L1 cache, followed by L2 cache, etc., and then RAM. As a rule, lower latency implies less storage capacity. The fact that information travels no faster than the speed of light within computer hardware explains this. This is known as the principle of memory locality: memory which is closer to the point of processing is faster to access.

The principle of memory locality has a very pronounced effect on the performance of code running on hardware, since memory access latency is often much higher than processing latency. In the context of SNARK proving, the principle of memory locality does not apply in the same way. There is still a general tendency that accessing smaller memories has less cost, but this is less pronounced than in the case of CPUs. SNARKs work with immutable, timeless mathematical relations, whereas hardware works with chains of cause and effect. Since information does not travel through space and time in the relations of SNARK proofs, there is no principle of memory locality for SNARKs in the physical sense having to do with the speed of information travel. In SNARK proving, there is a general tendency that updating smaller memories can be done by committing to less information, and this can make the costs of accessing smaller memories less.

In common with all modern CPU architectures, the architecture of RISC-V uses general-purpose registers to store inputs and outputs of logical and arithmetic operations. This results in a need to move data between RAM and registers, particularly at function call boundaries, where the contents of registers must be saved and restored by the caller and/or the callee (according to the calling convention). Compared to Valida, code generation for RISC-V will tend to emit more opcodes that deal with loading and storing data.

The use of general-purpose registers has a cost in terms of complexity of generated code. In the context of a CPU architecture, general-purpose registers have a benefit which outweighs the cost. On a typical program, the processor runs much faster than it would if it did not have general-purpose registers. On the other hand, in SNARK proving, there is not a benefit that outweighs the cost for having general-purpose registers. As Lita, we believe that this is a major reason why according to Lita's testing \cite{litabench} and the feedback Lita receives from users, Valida offers faster proving compared to zkVMs using RISC-V. 

\section{Scope}

This document specifies the basic Valida instruction set architecture (ISA). This is the ISA supported by the default machine in Lita's implementation of Valida. 

Valida is a Harvard architecture, with three separate address spaces holding the program code, the program data, and the RAM. The program code address space is executable, but not readable or writable. The program data address space is not executable, readable, or writable, and its values are simply initialized into RAM prior to program startup. The RAM address space is readable and writable, but not executable.

Valida is a 32-bit, little endian architecture. It has two special purpose registers: the frame pointer (FP) and the program counter (PC). Most opcodes directly address stack operands, specified by a fixed offset from FP. This is in lieu of having general purpose registers. All interaction of a Valida program with its environment is via a sequentially readable input tape and a sequentially writable output tape.

Valida is a modular ISA. Each chip in Valida introduces zero or more opcodes. By specifying a set of chips, one gets a Valida ISA. This document covers the basic Valida ISA, which is the ISA given by the set of chips in Figure~\ref{fig:basic-machine}, and enumerated below:

\begin{figure}
\includegraphics[width=\textwidth]{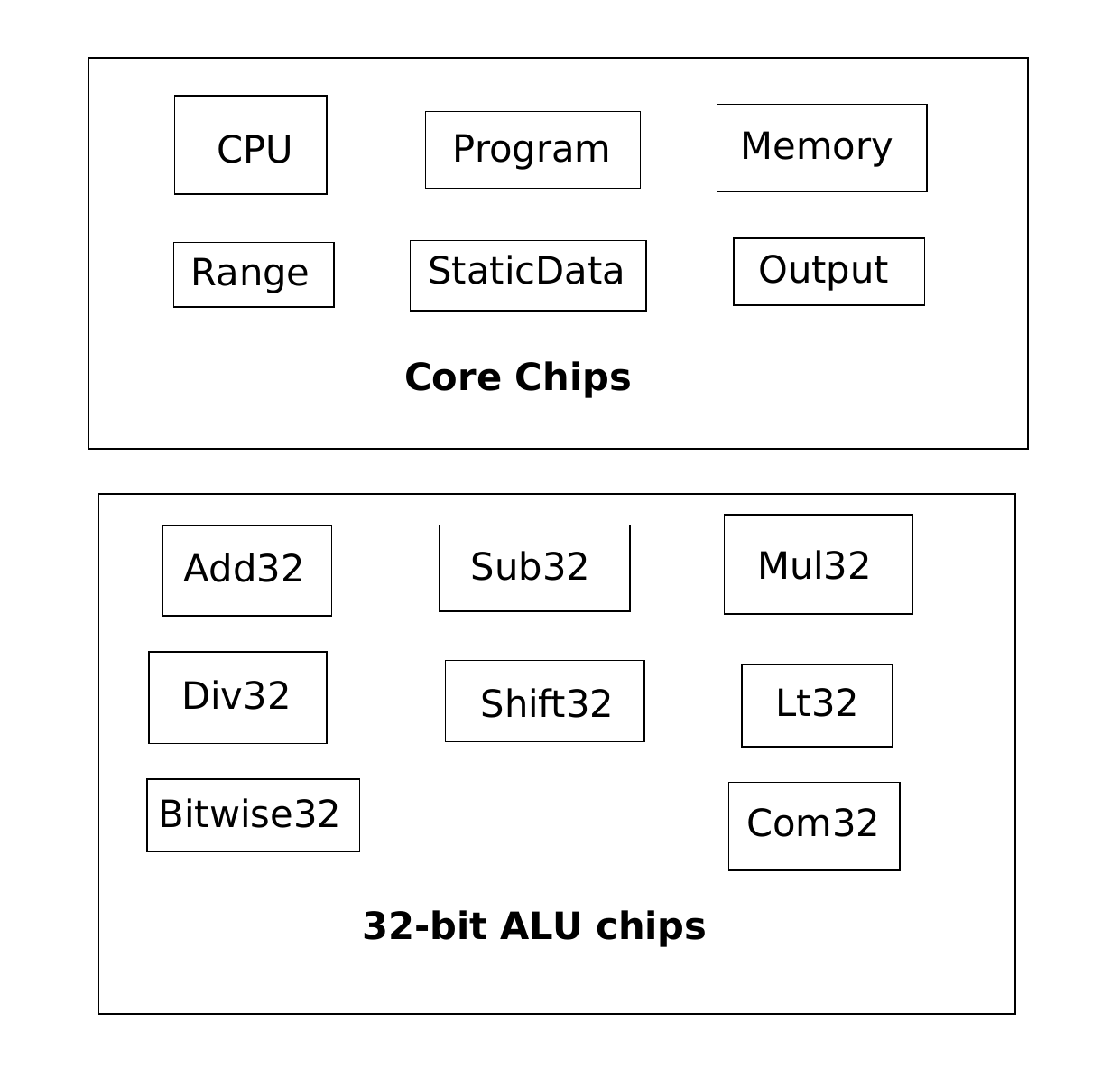}
\caption{The chips in the basic Valida machine.}
\label{fig:basic-machine}
\end{figure}

\begin{figure}
\begin{tabular}{llllllll}
\textbf{Op} & \textbf{A} & \textbf{B} & \textbf{C} & \textbf{Description} \\
Store32 & $A$ & $B$ & & Store the word at $fp+B$ to the word-aligned address at $fp+A$. \\
StoreU8 & $A$ & $B$ & & Store the LSB of the byte-valued word at $fp+B$ to the address at $fp+A$. \\
Load32 & $A$ & $B$ & & Store to $fp+A$ the word at the word-aligned address at $fp+B$. \\
LoadU8 & $A$ & $B$ & & Store to the word at $fp+A$ the unsigned byte at the address at $fp+B$. \\
LoadS8 & $A$ & $B$ & & Store to the word at $fp+A$ the sign-extended byte at the address at $fp+B$. \\
Jal & $A$ & $B$ & $C$ & Set $pc$ to $B$, store $pc$ to $fp+A$, and add $C$ to $fp$. \\
Jalv & $A$ & $B$ & $C$ & Set $pc$ to the word at $fp+B$, store $fp$ at $fp+A$, and set $fp$ to the word at $fp+A$. \\
Beq & $A$ & $B$ & $C$ & Set $pc$ to $A$ if the word at $fp+B$ is the word at $fp+C$. \\
Beqi & $A$ & $B$ & $C$ & Set $pc$ to $A$ if the word at $fp+B$ is $C$. \\
Bne & $A$ & $B$ & $C$ & Set $pc$ to $A$ if the word at $fp+B$ is not the word at $fp+C$. \\
Bnei & $A$ & $B$ & $C$ & Set $pc$ to $A$ if the word at $fp+B$ is not $C$. \\
Imm32 & $A$ & $B$ & & Set the word at $fp+A$ to $B$. \\
ReadAdvice & $A$ & & & Set the word at $fp+A$ to the next byte of the input. \\
Stop & & & & Halt the program successfully. \\
LoadFp & $A$ & $B$ & & Store $fp+B$ to the word at $fp+A$. \\
Write & $A$ & & & Append the LSB of the word at $fp+A$ to the output. \\
Add & $A$ & $B$ & $C$ & Set the word at $fp+A$ to the word at $fp+B$ plus the word at $fp+C$. \\
Addi & $A$ & $B$ & $C$ & Set the word at $fp+A$ to the the word at $fp+B$ plus $C$. \\
Addc & $A$ & $B$ & $C$ & Set the word at $fp+A$ to the sum carry of the words at $fp+B$ and $fp+C$. \\
Addci & $A$ & $B$ & $C$ & Set the word at $fp+A$ to the sum carry of the word at $fp+B$ and $C$. \\
Sub & $A$ & $B$ & $C$ & Set the word at $fp+A$ to the word at $fp+B$ minus the word at $fp+C$. \\
Subi & $A$ & $B$ & $C$ & Set the word at $fp+A$ to the the word at $fp+B$ minus $C$. \\
Isub &  $A$ & $B$ & $C$ & Set the word at $fp+A$ to $B$ minus the word at $fp+C$. \\
Subb & $A$ & $B$ & $C$ & Set the word at $fp+A$ to the borrow of the words at $fp+B$ minus $fp+C$. \\
Subbi & $A$ & $B$ & $C$ & Set the word at $fp+A$ to the borrow of the word at $fp+B$ minus $C$. \\
Isubb &  $A$ & $B$ & $C$ & Set the word at $fp+A$ to the borrow of $B$ minus the word at $fp+C$. \\
Mul & $A$ & $B$ & $C$ & Set the word at $fp+A$ to the product of the words at $fp+B$ and $fp+C$. \\
Muli & $A$ & $B$ & $C$ & Set the word at $fp+A$ to the product LSBs of the word at $fp+B$ and $C$. \\
Mulhs & $A$ & $B$ & $C$ & Store the signed product MSBs of the words at $fp+B$ and $fp+C$ at $fp+A$. \\
Mulhsi & $A$ & $B$ & $C$ & Store the signed product MSBs of the word at $fp+B$ and $C$ at $fp+A$. \\
Mulhu & $A$ & $B$ & $C$ & Store the unsigned product MSBs of the words at $fp+B$ and $fp+C$ at $fp+A$. \\
Mulhui & $A$ & $B$ & $C$ & Store the unsigned product MSBs of the word at $fp+B$ and $C$ at $fp+A$. \\
Div & $A$ & $B$ & $C$ & Set the word at $fp+A$ to the quotient of the words at $fp+B$ over $fp+C$. \\
Divi & $A$ & $B$ & $C$ & Set the word at $fp+A$ to the quotient of the word at $fp+B$ over $C$. \\
Sdiv & $A$ & $B$ & $C$ & Set the word at $fp+A$ to the signed quotient of the words at $fp+B$ over $fp+C$. \\
Sdivi & $A$ & $B$ & $C$ & Set the word at $fp+A$ to the signed quotient of the word at $fp+B$ over $C$. \\
Shl & $A$ & $B$ & $C$ & Set the word at $fp+A$ to the word at $fp+B$ bit-shift left the word at $fp+C$. \\
Shli & $A$ & $B$ & $C$ & Set the word at $fp+A$ to the word at $fp+B$ bit-shift left $C$. \\
Ishl & $A$ & $B$ & $C$ & Set the word at $fp+A$ to $B$ bit-shift left the word at $fp+C$. \\
Shr & $A$ & $B$ & $C$ & Set the word at $fp+A$ to the word at $fp+B$ bit-shift right the word at $fp+C$. \\
Shri & $A$ & $B$ & $C$ & Set the word at $fp+A$ to the word at $fp+B$ bit-shift right $C$. \\
Ishr & $A$ & $B$ & $C$ & Set the word at $fp+A$ to $B$ bit-shift right the word at $fp+C$. \\
Sra & $A$ & $B$ & $C$ & Set the word at $fp+A$ to the words at $fp+B$ arithmetic shift right $fp+C$. \\
Srai & $A$ & $B$ & $C$ & Set the word at $fp+A$ to the word at $fp+B$ arithmetic shift right $C$. \\
Isra & $A$ & $B$ & $C$ & Set the word at $fp+A$ to $B$ arithmetic shift right $C$ the word at $fp+C$. \\
\end{tabular}
\caption{The Valida opcodes (1 of 2).}
\label{tbl:opcodes1}
\end{figure}

\begin{figure}
\begin{tabular}{llllllll}
\textbf{Op} & \textbf{A} & \textbf{B} & \textbf{C} & \textbf{Description} \\
Lt & $A$ & $B$ & $C$ & Set the word at $fp+A$ to say if the word at $fp+B$ is less than the word at $fp+C$. \\
Lti & $A$ & $B$ & $C$ & Set the word at $fp+A$ to say if the word at $fp+B$ is less than $C$. \\
Ilt & $A$ & $B$ & $C$ & Set the word at $fp+A$ to say if the word at $fp+B$ is less than $C$. \\
Lte & $A$ & $B$ & $C$ & Set the word at $fp+A$ to say if the word at $fp+C$ is greater than the word at $fp+B$. \\
Ltei & $A$ & $B$ & $C$ & Set the word at $fp+A$ to say if $C$ is greater than the word at $fp+B$. \\
Ilte & $A$ & $B$ & $C$ & Set the word at $fp+A$ to say if the word at $fp+C$ is greater than $B$. \\
Slt & $A$ & $B$ & $C$ & Set the word at $fp+A$ to say if the word at $fp+B$ is less than the word at $fp+C$. \\
    &     &     &     & (Signed variant.) \\
Slti & $A$ & $B$ & $C$ & Set the word at $fp+A$ to say if the word at $fp+B$ is less than $C$. (Signed variant.) \\
Islt & $A$ & $B$ & $C$ & Set the word at $fp+A$ to say if $B$ is less than the word at $fp+C$. (Signed variant.) \\
Slte & $A$ & $B$ & $C$ & Set the word at $fp+A$ to say if the word at $fp+C$ is greater than the word at $fp+B$. \\
    &     &     &     & (Signed variant.) \\
Sltei & $A$ & $B$ & $C$ & Set the word at $fp+A$ to say if $C$ is greater than the word at $fp+B$. (Signed variant.) \\
Eq & $A$ & $B$ & $C$ & Set the word at $fp+A$ to say if the word at $fp+B$ is the word at $fp+C$. \\
Eqi & $A$ & $B$ & $C$ &  Set the word at $fp+A$ to say if the word at $fp+B$ is $C$. \\
Ne & $A$ & $B$ & $C$ & Set the word at $fp+A$ to say if the word at $fp+B$ is not the word at $fp+C$. \\
Nei & $A$ & $B$ & $C$ & Set the word at $fp+A$ to say if the word at $fp+B$ is not $C$. \\
And & $A$ & $B$ & $C$ & Set the word at $fp+A$ to the bitwise conjunction of the words at $fp+B$ and $fp+C$. \\
Andi & $A$ & $B$ & $C$ & Set the word at $fp+A$ to the bitwise conjunction of the word at $fp+B$ and $C$. \\
Or & $A$ & $B$ & $C$ & Set the word at $fp+A$ to the inclusive bit disjunction of the words at $fp+B$ and $fp+C$. \\
Ori & $A$ & $B$ & $C$ & Set the word at $fp+A$ to the inclusive bi tdisjunction of the word at $fp+B$ and $C$. \\
Xor & $A$ & $B$ & $C$ & Set the word at $fp+A$ to the exclusive bit disjunction of the words at $fp+B$ and $fp+C$. \\
Xori & $A$ & $B$ & $C$ & Set the word at $fp+A$ to the exclusive bit disjunction of the word at $fp+B$ and $C$. \\
\end{tabular}
\caption{The Valida opcodes (2 of 2).}
\label{tbl:opcodes2}
\end{figure}

\begin{enumerate}
\item \textbf{The CPU chip} provides core opcodes for memory access, flow control, loading constants, I/O, and reading special-purpose registers.
\item \textbf{The Program chip} provides no opcodes, but stores the program code. The program chip and the static data chip together store the program.
\item \textbf{The StaticData chip} provides no opcodes, but stores the program data, i.e., the initial static data values that are loaded into RAM prior to execution.
\item \textbf{The Memory chip} provides no opcodes, but stores the RAM access trace, which specifies all of the RAM reads and writes, including their results.
\item \textbf{The Range chip} provides no opcodes, but is used by other chips to check that values are in the range 0 to 255, inclusive.
\item \textbf{The Output chip} provides no opcodes, but stores the output trace, which specifies all of the data written by the program to the output tape.
\item \textbf{The Add32 chip} provides opcodes for 32-bit addition.
\item \textbf{The Sub32 chip} provides opcodes for 32-bit subtraction.
\item \textbf{The Mul32 chip} provides opcodes for 32-bit multiplication.
\item \textbf{The Div32 chip} provides opcodes for 32-bit division.
\item \textbf{The Shift32 chip} provides opcodes for 32-bit arithmetic shifts and logical shifts.
\item \textbf{The Lt32 chip} provides opcodes for 32-bit inequality comparisons.
\item \textbf{The Com32 chip} provides opcodes for 32-bit equality comparisons.
\end{enumerate}

\section{Notation}

This spec uses the following standard mathematical notations.

Sets are unordered collections of objects.
Sets are extensional, i.e., two sets are equal if and only if they have all the same elements.
$x \in S$ means that $S$ is a set and $x$ is an element of $S$.

$\emptyset$ denotes the empty set. Finite sets can be denoted by enumerating their elements in between curly braces, e.g.: $\{0,1,2\}$. Set builder notation can be used to denote a subset of a set, the subset of all elements satisyfing a condition. For example, the set

\begin{equation}
\{x \in \{0,1,2,3\}\ |\ x < 1\}
\end{equation}

is pronounced ``the set of $x$ such that $x$ is in $\{0,1,2,3\}$ and $x$ is less than one,'' and is equal to $\{0\}$. Although it is not always well defined, set builder notation can also be used without specifying a set that the set being built is a subset of, but just specifying the condition that elements must satisfy, as in Equation~\ref{eq:cartesian-product} below.

The Cartesian product of sets $A$ and $B$ is denoted $A \times B$. A Cartesian product is a set of ordered pairs:

\begin{equation}
A \times B = \{(a, b)\ |\ a \in A\ \text{and}\ b \in B\}.
\label{eq:cartesian-product}
\end{equation}

A function $f : A \to B$ is a subset of the Cartesian product $A \times B$ of sets $A$ and $B$, $f \subseteq A \times B$, such that for every $a \in A$, there is a unique $b \in B$ such that $(a, b) \in f$. If $f : A \to B$ is a function, and $x \in A$, then $f(x)$ denotes the unique $b \in B$ such that $(x, b) \in f$.

The inverse of a function $f : A \to B$, or more generally a set $f \subseteq A \times B$, is denoted $f^{-1}$ and is defined as the set:
\begin{equation}
f^{-1} := \{(b, a) \in B \times A\ |\ (a, b) \in f\}.
\end{equation}
The inverse $f^{-1}$ is a subset of $B \times A$. $f^{-1}$ is sometimes not a function. In the case where $f^{-1}$ is a function, $f$ is called a bijection. A bijection is also called an isomorphism of sets.

For any given $f : A \times B$ and $y \in B$, if there is a unique $x \in A$ such that $f(x) = y$, then $f^{-1}(y)$ denotes that $x$. If there not a unique $x \in B$ such that $f(x) = y$, then the notation $f^{-1}(y)$ does not have a denotation.

The generic projection functions $\pi_1 : A \times B \to A$ and $\pi_2 : A \times B \to B$ are defined as follows.
\begin{equation}
\pi_1((a,b)) := a.
\end{equation}
\begin{equation}
\pi_2((a,b)) := b.
\end{equation}

The Cartesian product and its projection functions can be extended to triples, quadruples, and so forth, by for example, by defining $A \times B \times C$ as $(A \times B) \times C$.

The union of sets $A$ and $B$ is denoted $A \cup B$:
\begin{equation}
A \cup B := \{x\ |\ x \in A\ \text{or}\ x \in B\}.
\end{equation}

The difference of sets $A$ and $B$ is denoted $A \setminus B$:
\begin{equation}
A \setminus B := \{x \in A\ |\ x \notin B\}.
\end{equation}

Natural numbers (i.e., non-negative integers) are represented as von Neumann ordinals, i.e., as the set of smaller natural numbers. For example, the natural number 0 is represented as the empty set, also denoted $\emptyset$. The number 1 is represented as $\{0\}$. The number 2 is represented as $\{0,1\}$. 

In general, the natural number $n$ is represented as the set $\{0, ..., n-1\}$. So for example, ``$2^{32}$'' denotes the set of 32-bit non-negative integers. The set of all natural numbers is denoted $\mathbb{N}$ is defined as the set of all finite von Neumann ordinals: $\{0, 1, 2, ...\}$.

The Kleene star notation is used to denote sets of strings. If $S$ is a set, then $S^*$ (pronounced ``$S$ star'' or ``the Kleene star of $S$'') is the set of (finite length) strings of elements of $S$. If $a_0, ..., a_n$ are elements of $S$, then $(a_0, ..., a_n)$ denotes the string of those elements. $()$ denotes the empty string. If $s_0, s_1 \in S^*$, then let $s_0 +^* s_1$ denote the concatenation of $s_0$ and $s_1$.

For any set $S$, let
\begin{equation}
\text{head} : (S^* \setminus \{()\}) \to S
\end{equation}
be the function which takes the first element of a non-empty string. Let:
\begin{equation}
\text{tail} : (S^* \setminus \{()\}) \to S^*
\end{equation}
be the function which takes everything but the first element of a non-empty string. Let:
\begin{equation}
\text{tail}' : S^* \to S^*
\end{equation}
be the function which is like tail except that it is defined to return the empty string () when given the empty string as input.

If $S$ and $T$ are sets, then $S \to T$ denotes the set of functions with domain $S$ and codomain $T$. The notation $f : S \to T$ means that $f$ is a function in $S \to T$. Depending on the context, $S \to T$ can also denote the set of partial functions from $S$ to $T$.

An indexed set of sets such as $\{A_i\}_{i \in S}$ is in other words a function $S \to \text{Set}$ which maps each element $i \in S$ to a set $A_i$. Given an indexed set of sets $\{A_i\}_{i \in S}$, its Cartesian product is denoted by $\prod_{i \in S} A_i$. This is a generalization of the Cartesian product of two sets. The indexed product $\prod_{i \in S} A_i$ is in other words the dependent function space $(i : S) \to A_i$, which can be defined using set builder notation:
\begin{equation}
\prod_{i \in S} A_i := \{f\ |\ f\ \text{is a function with domain}\ S\ \text{and for all}\ i \in S,\ f(i) \in A_i\}.
\end{equation}

If $f : A \to B$ is a function, and $(a,b) \in A \times B$, then $f[a \mapsto b]$ denotes the function in $A \to B$ which is identical to $f$ except that it maps $a$ to $b$ (replacing any mapping from $a$ that is in $f$):
\begin{equation}
f[a \mapsto b] := (f \setminus \{(a,b')\ |\ b' \in B\}) \cup \{(a,b)\}.
\end{equation}
If $(a_1,b_1), ..., (a_n, b_n)$ are in $A \times B$, then:
\begin{equation}
f[a_1 \mapsto b_1, ..., a_n \mapsto b_n] := f[a_1 \mapsto b_1]\cdots[a_n \mapsto b_n].
\end{equation}

The union of an indexed set of sets $\{A_i\}_{i \in S}$ can be written $\bigcup_{i \in S} A_i$ and defined as:
\begin{equation}
\bigcup_{i \in S} A_i := \{a |\ \text{there exists}\ i \in S\ \text{such that}\ a \in A_i\}.
\end{equation}

Equality definition is denoted by $:=$. For example, to say that $A$ is 2 by definition, write $A := 2$.

Isomorphism is denoted by $\cong$. To say that an isomorphism exists between $A$ and $B$, write $A \cong B$.

``$\mathbb{Z}_n$'' denotes the ring of integers modulo $n$. Addition in $\mathbb{Z}_n$ is denoted $+_{\mathbb{Z}_n}$. Multiplication in the ring $\mathbb{Z}_n$ is denoted $\times_{\mathbb{Z}_n}$. The set of elements of $\mathbb{Z}_n$ can conveniently be defined as the von Neumann ordinal $n$. The addition and multiplication operations are defined in the usual way for integers mod $n$.

``$\mathbb{Z}$'' denotes the ring of integers. Addition in the ring of integers $\mathbb{Z}$ is denoted $+_{\mathbb{Z}}$, and the corresponding subtraction is denoted $-_{\mathbb{Z}}$. Multiplication in the ring of integers is denoted $\times_{\mathbb{Z}}$. The set of elements of the ring of integers is defined as the set:
\begin{equation}
\{0\} \cup (\{-,+\} \times (\mathbb{N} \setminus \{0\})).
\end{equation}

Division in the ring of integers is denoted $\div_{\mathbb{Z}}$. Division is only a partial operation on the ring of integers. It is undefined when the denominator is zero or does not divide the numerator.

For any $x \in \mathbb{Z}$, $|x|$ denotes the absolute value of $x$, which is defined as follows:
\begin{equation}
|x| := \begin{cases}
x & \text{if}\ 0 \leq x, \\
-x & \text{if}\ x < 0.
\end{cases}
\end{equation}

\section{Outline}

This spec consists of the following basic parts:
\begin{enumerate}
\item The \textbf{program model definition} defines a set $\Pi$ whose elements represent all possible basic Valida programs.
\item The \textbf{state model definition} defines a set $\Sigma$ whose elements represent all possible states of a basic Valida program execution.
\item The \textbf{initial state definition} defines a function $\iota : \Pi \times (2^{32})^* \to \Sigma$ which maps a basic Valida program, and a string of 32-bit values representing the contents of the input tape, to the resulting initial state of the execution of that program with that input.
\item The \textbf{transition function definition} defines a function $\tau : \Sigma \to \Sigma$. A single application of $\tau$ maps a state to the resulting state after executing the next instruction. $\tau$ maps the state of a halted program onto the same state.
\item The \textbf{result function definition} defines a function
\begin{equation}
  \rho : \Sigma \to (\{\text{Halted}, \text{NotHalted}\} \times (2^{32})^*).
\end{equation}
This function maps a program execution state to the result of that execution so far, including whether it has halted or not, and the contents of the output tape so far.
\end{enumerate}

\section{Program, state model, initial state, and result function definitions}

A Valida program consists of its code and its data, plus its correct initial values for PC and FP:
\begin{equation}
\label{eq:program}
\Pi := \text{ProgramCode} \times \text{ProgramData} \times \text{PC} \times \text{FP}.
\end{equation}

Let $4(2^{30})$ denote the set of elements of $2^{32}$ which are multiples of 4, i.e.:
\begin{equation}
4(2^{30}) := \{4 \times i\ |\ i \in 2^{30}\}.
\end{equation}

Similarly, let $24(178956970)$ denote the set of elements of $2^{32}$ which are multiplies of 24, i.e.:
\begin{equation}
24(178956970) := \{24 \times i\ |\ i \in 178956970\}.
\end{equation}

\begin{equation}
\text{PC} := 24(178956970).
\end{equation}

\begin{equation}
\text{FP} := 4(2^{30}).
\end{equation}

The program data lives in a 32-bit, byte-indexed address space, and can be defined as the following set of partial functions:
\begin{equation}
\text{ProgramData} := 2^{32} \to 2^8.
\end{equation}

The program code lives in an instruction-indexed address space, with as many addresses as possible, subject to the byte-addressed program addresses being 32-bit. In the Valida state model, instructions are represented in decoded form. The program code model can be defined as the following set of partial functions:
\begin{equation}
\text{ProgramCode} := \text{CodeAddress} \to \text{Instruction}.
\end{equation}
\begin{equation}
\text{CodeAddress} := 178956970.
\end{equation}
\begin{equation}
\text{Instruction} := \bigcup_{i \in \text{Chips}} \text{Instruction}_i.
\end{equation}

Here is the set of chips:

\begin{equation}
\begin{array}{rl}
\text{Chips} := \{ & \text{CPU, Program, Memory, Range, StaticData, Output,} \\
                   & \text{Add32, Sub32, Mul32, Div32, Shift32, Lt32, Com32, Bitwise32}\ \}.
\end{array}
\end{equation}

Here are the instruction set definitions for each chip. For the purposes of this spec, the opcodes $\text{Opcode}_i$ can be any objects, as long as they are all distinct from each other.

\begin{equation}
\begin{array}{rcl}
\text{Instruction}_{\text{CPU}} &:=& \text{Store32} \cup \text{StoreU8} \cup \text{Load32} \cup \text{LoadU8} \\
  &\cup& \text{LoadS8} \cup \text{Jal} \cup \text{Jalv} \cup \text{Beq} \cup \text{Beqi} \cup \text{Bne} \cup \text{Bnei} \cup \text{Imm32} \\
  &\cup& \text{ReadAdvice} \cup \text{Stop} \cup \text{LoadFp}.
\end{array}
\end{equation}

\begin{equation}
\text{Store32} := \{\text{Opcode}_{\text{Store32}}\} \times 4(2^{30}) \times 4(2^{30}).
\end{equation}

\begin{equation}
\text{StoreU8} := \{\text{Opcode}_{\text{StoreU8}}\} \times 4(2^{30}) \times 4(2^{30}).
\end{equation}

\begin{equation}
\text{Load32} := \{\text{Opcode}_{\text{Load32}}\} \times 4(2^{30}) \times 4(2^{30}).
\end{equation}

\begin{equation}
\text{LoadU8} := \{\text{Opcode}_{\text{LoadU8}}\} \times 4(2^{30}) \times 4(2^{30}).
\end{equation}

\begin{equation}
\text{LoadS8} := \{\text{Opcode}_{\text{LoadS8}}\} \times 2^{32} \times 2^{32}.
\end{equation}

\begin{equation}
\text{Jal} := \{\text{Opcode}_{\text{Jal}}\} \times 4(2^{30}) \times 24(178956970) \times 4(2^{30}).
\end{equation}

\begin{equation}
\text{Jalv} := \{\text{Opcode}_{\text{Jalv}}\} \times 4(2^{30}) \times 4(2^{30}) \times 4(2^{30}).
\end{equation}

\begin{equation}
\text{Beq} := \{\text{Opcode}_{\text{Beq}}\} \times 24(178956970) \times 4(2^{30}) \times 4(2^{30}).
\end{equation}

\begin{equation}
\text{Beqi} := \{\text{Opcode}_{\text{Beqi}}\} \times 24(178956970) \times 4(2^{30}) \times 2^{32}.
\end{equation}

\begin{equation}
\text{Bne} := \{\text{Opcode}_{\text{Bne}}\} \times 24(178956970) \times 4(2^{30}) \times 4(2^{30}).
\end{equation}

\begin{equation}
\text{Bnei} := \{\text{Opcode}_{\text{Bnei}}\} \times 24(178956970) \times 4(2^{30}) \times 2^{32}.
\end{equation}

\begin{equation}
\text{Imm32} := \{\text{Opcode}_{\text{Imm32}}\} \times 2^{32} \times 2^{32}.
\end{equation}

\begin{equation}
\text{ReadAdvice} := \{\text{Opcode}_{\text{ReadAdvice}}\} \times 4(2^{30}).
\end{equation}

\begin{equation}
\text{Stop} := \{\text{Opcode}_{\text{Stop}}\}.
\end{equation}

\begin{equation}
\text{LoadFp} := \{\text{Opcode}_{\text{LoadFp}}\} \times 4(2^{30}) \times 2^{32}.
\end{equation}

\begin{equation}
\text{Instruction}_{\text{Program}} := \emptyset.
\end{equation}

\begin{equation}
\text{Instruction}_{\text{Memory}} := \emptyset.
\end{equation}

\begin{equation}
\text{Instruction}_{\text{Range}} := \emptyset.
\end{equation}

\begin{equation}
\text{Instruction}_{\text{Output}} := \text{Write}.
\end{equation}

\begin{equation}
\text{Write} := \{\text{Opcode}_\text{Write}\} \times 2^{32}.
\end{equation}

\begin{equation}
\text{Instruction}_{\text{Add32}} := \text{Add} \cup \text{Addi} \cup \text{Addc} \cup \text{Addci}.
\end{equation}

\begin{equation}
\text{Instruction}_{\text{Sub32}} := \text{Sub} \cup \text{Subi} \cup \text{Isub} \cup \text{Subb} \cup \text{Subbi} \cup \text{Isubb}.
\end{equation}

\begin{equation}
\text{Instruction}_{\text{Mul32}} := \text{Mul} \cup \text{Muli} \cup \text{Mulhs} \cup \text{Mulhsi} \cup \text{Mulhu} \cup \text{Mulhui}.
\end{equation}

\begin{equation}
\text{Instruction}_{\text{Div32}} := \text{Div} \cup \text{Divi} \cup \text{Sdiv} \cup \text{Sdivi}.
\end{equation}

\begin{equation}
\text{Instruction}_{\text{Shift32}} := \text{Shl} \cup \text{Shli} \cup \text{Ishl} \cup \text{Shr} \cup \text{Shri} \cup \text{Ishr} \cup \text{Sra} \cup \text{Srai} \cup \text{Isra}.
\end{equation}

\begin{equation}
\text{Instruction}_{\text{Lt32}} := \text{Lt} \cup \text{Lti} \cup \text{Ilt} \cup \text{Lte} \cup \text{Ltei} \cup \text{Ilte} \cup \text{Slt} \cup \text{Slti} \cup \text{Islt} \cup \text{Slte} \cup \text{Islte} \cup \text{Sltei}.
\end{equation}

\begin{equation}
\text{Instruction}_{\text{Com32}} := \text{Eq} \cup \text{Eqi} \cup \text{Ne} \cup \text{Nei}.
\end{equation}

\begin{equation}
\text{Instruction}_{\text{Bitwise32}} := \text{And} \cup \text{Andi} \cup \text{Or} \cup \text{Ori} \cup \text{Xor} \cup \text{Xori}.
\end{equation}

For all $i \in \{\text{Add, Addc, Sub, Subb, Mul, Mulhs, Mulhu, Div, Sdiv, Shl, Shr, Sra, Lt, Lte, Slt, Slte, Eq}$, $\text{Ne, And, Or, Xor}\}$:

\begin{equation}
\label{eq:generic-binary-op-vv}
i := \{\text{Opcode}_i\} \times 4(2^{30}) \times 4(2^{30}) \times 4(2^{30}).
\end{equation}

For all $i \in \{\text{Addi, Addci, Subi, Subbi, Muli, Mulhsi, Mulhui, Divi, Sdivi, Shli, Shri, Srai, Lti, Ltei, Slti}$, $\text{Sltei, Eqi, Nei, Andi, Ori, Xori}\}$:

\begin{equation}
\label{eq:generic-binary-op-vi}
i := \{\text{Opcode}_i\} \times 4(2^{30}) \times 4(2^{30}) \times 2^{32}.
\end{equation}

For all $i \in \{\text{Isub, Isubb, Ishl, Ishr, Isra, Ilt, Ilte, Islt, Islte}\}$:

\begin{equation}
\label{eq:generic-binary-op-iv}
i := \{\text{Opcode}_i\} \times 4(2^{30}) \times 2^{32} \times 4(2^{30}).
\end{equation}

The Valida state model can be defined as simply the Cartesian product of the state models of all chips:

\begin{equation}
\Sigma := \prod_{i \in \text{Chips}} \Sigma_i.
\end{equation}

For each $i \in \text{Chips}$, let $\pi_i : \Sigma \to \Sigma_i$ denote the projection function which maps a machine state $s \in \Sigma$ to the state of chip $i$ in $s$.

Most of the chips are stateless. This means that their state models contain no information. A stateless chip's state can be modeled as a set with one element, such as the von Neumann ordinal $1 = \{0\}$. These stateless chips' state models can safely be omitted from the basic Valida state model, without changing the isomorphism class.

The following chips are stateful: CPU, Program, StaticData, Memory, and Output. All other chips are stateless. Program and StaticData have immutable state, meaning that their states do not change during a program execution. CPU, Memory, and Output have mutable state, meaning that their states can change during a program execution.

The Valida state model can be explicitly represented this way:

\begin{equation}
\Sigma \cong \Sigma_{\text{CPU}} \times \Sigma_{\text{Program}} \times \Sigma_{\text{StaticData}} \times \Sigma_{\text{Memory}} \times \Sigma_{\text{Output}}.
\end{equation}

What follows are each of the stateful chips' state model definitions, and the definitions they depend on.

The CPU chip's state consists of the special purpose register states, plus the remaining contents of the input tape, and a boolean value indicating if the program has halted or not.

\begin{equation}
\Sigma_{\text{CPU}} := \text{PC} \times \text{FP} \times \text{UnconsumedInput} \times \{\text{Halted, NotHalted}\}.
\end{equation}

\begin{equation}
\text{UnconsumedInput} := (2^{32})^*.
\end{equation}

Let there be the following projection functions:

\begin{equation}
\pi_{\text{PC}} : \Sigma_{\text{CPU}} \to \text{PC}.
\end{equation}

\begin{equation}
\pi_{\text{FP}} : \Sigma_{\text{CPU}} \to \text{FP}.
\end{equation}

\begin{equation}
\pi_{\text{UnconsumedInput}} : \Sigma_{\text{CPU}} \to \text{UnconsumedInput}.
\end{equation}

\begin{equation}
\pi_{\text{Halting}} : \Sigma_{\text{CPU}} \to \{\text{Halted}, \text{NotHalted}\}.
\end{equation}

The Program chip's state is simply the program code, as in the first coordinate of $\Pi$. It can be modeled as the following set of partial functions:

\begin{equation}
\Sigma_{\text{Program}} := \text{CodeAddress} \to \text{Instruction}.
\end{equation}

The StaticData chip's state is the program data, as in the second coordinate of $\Pi$ (Equation~\ref{eq:program}). It can be modeled as the following set of partial functions:

\begin{equation}
\Sigma_{\text{StaticData}} := 2^{32} \to 2^8.
\end{equation}

The Memory chip's state is the contents of RAM. It can be modeled as the same set of partial functions:

\begin{equation}
\Sigma_{\text{Memory}} := 2^{32} \to 2^8.
\end{equation}

Memory is stored in little endian order. The following function can be used to load a word from memory:
\begin{equation}
\text{load} : \Sigma \times 2^{32} \to 2^{32}.
\end{equation}
\begin{equation}
\text{load}(s, a) := \sum_{i=0}^3 2^{8i} \times_{\mathbb{Z}_{2^{32}}} \pi_{\text{Memory}}(s)(a +_{\mathbb{Z}_{2^{32}}} i).
\end{equation}

The following function can be used to store a word to memory:
\begin{equation}
\text{store} : \Sigma \times 4(2^{30}) \times 2^{32} \to \Sigma_{\text{Memory}}.
\end{equation}
\begin{equation}
\text{store}(s, a, \sum_{i=0}^3 2^{8i}x_i) := \pi_{\text{Memory}}(s)[a \mapsto x_0, a+1 \mapsto x_1, a+2 \mapsto x_2, a+3 \mapsto x_3].
\end{equation}

The Output chip's state is the contents of the output tape:

\begin{equation}
\Sigma_{\text{Output}} := (2^{32})^*.
\end{equation}

That completes the state model definition.

Here is the initial state definition:

\begin{equation}
\iota(c,d,pc_0,fp_0) = \left\{ \left (i,
    \begin{cases}
      (0, pc_0 \div_{\mathbb{Z}} 24, fp_0, i, \text{NotHalted})) & \text{if}\ i = \text{CPU}, \\
      c & \text{if}\ i = \text{Program}, \\
      d & \text{if}\ i = \text{StaticData}, \\
      d & \text{if}\ i = \text{RAM}, \\
      () & \text{if}\ i = \text{Output}, \\
      0 & \text{otherwise}
    \end{cases}
  \right )\ |\ i \in \text{Chips} \right\}.
\end{equation}

Here is the result function definition:

\begin{equation}
\rho(s) = (\pi_{\text{Halting}}(\pi_{\text{CPU}}(s)), \pi_{\text{Output}}(s)).
\end{equation}

\section{Preliminaries for transition function definition}

Let there be the following projection functions:
\begin{equation}
\pi_{u32} : 2^{32} \to \mathbb{Z}.
\end{equation}
\begin{equation}
\pi_{u32}(x) := \begin{cases}
0 & \text{if}\ x = 0, \\
+x & \text{otherwise}.
\end{cases}
\end{equation}
\begin{equation}
\pi_{s32} : 2^{32} \to \mathbb{Z}.
\end{equation}
\begin{equation}
\pi_{s32}(x) := \begin{cases}
0 & \text{if}\ x = 0, \\
+x & \text{if}\ 0 < x < 2^{31}, \\
-(2^{32} - x) & \text{otherwise}.
\end{cases}
\end{equation}

\begin{equation}
\pi_{u8} : 2^8 \to \mathbb{Z}.
\end{equation}
\begin{equation}
\pi_{u8}(x) := \begin{cases}
0 & \text{if}\ x = 0, \\
+x & \text{otherwise}.
\end{cases}
\end{equation}

\begin{equation}
\pi_{s8} : 2^8 \to \mathbb{Z}.
\end{equation}
\begin{equation}
\pi_{s8}(x) := \begin{cases}
0 & \text{if}\ x = 0, \\
+x & \text{if}\ x < 2^7, \\
-(2^8 - x) & \text{otherwise}.
\end{cases}
\end{equation}

Let there be the following functions, which truncate integers to 32 bits (signed or unsigned):
\begin{equation}
\text{trunc}_{u32} : \mathbb{Z} \to \mathbb{Z}.
\end{equation}
\begin{equation}
\text{trunc}_{u32}(x) := x +_{\mathbb{Z}} (2^{32} \times_{\mathbb{Z}} n),
\end{equation}
where $n \in \mathbb{Z}$ is the unique integer such that:
\begin{equation}
0 \leq x +_{\mathbb{Z}} (2^{32} \times_{\mathbb{Z}} n) < 2^{32}.
\end{equation}

\begin{equation}
\text{trunc}_{s32} : \mathbb{Z} \to \mathbb{Z}.
\end{equation}
\begin{equation}
\text{trunc}_{s32}(x) := x +_{\mathbb{Z}} (2^{32} \times_{\mathbb{Z}} n),
\end{equation}
where $n \in \mathbb{Z}$ is the unique integer such that:
\begin{equation}
-2^{31} \leq x +_{\mathbb{Z}} (2^{32} \times_{\mathbb{Z}} n) < 2^{31}.
\end{equation}

Let there be the following functions, which truncate integers to 8 or 5 bits (unsigned):
\begin{equation}
\text{trunc}_{u8} : \mathbb{Z} \to \mathbb{Z}.
\end{equation}
\begin{equation}
\text{trunc}_{u8}(x) := x +_{\mathbb{Z}} (2^8 \times_{\mathbb{Z}} n),
\end{equation}
where $n$ is the unique integer such that:
\begin{equation}
0 \leq x +_{\mathbb{Z}} (2^8 \times_{\mathbb{Z}} n) < 2^8.
\end{equation}

\begin{equation}
\text{trunc}_{u5} : \mathbb{Z} \to \mathbb{Z}.
\end{equation}
\begin{equation}
\text{trunc}_{u5}(x) := x +_{\mathbb{Z}} (2^5 \times_{\mathbb{Z}} n),
\end{equation}
where $n$ is the unique integer such that:
\begin{equation}
0 \leq x +_{\mathbb{Z}} (2^5 \times_{\mathbb{Z}} n) < 2^5.
\end{equation}

Let there be the following functions, which take the high bits of a result, discarding the lower 32 bits:
\begin{equation}
\text{high}_{u32} : \mathbb{Z} \to \mathbb{Z}.
\end{equation}
\begin{equation}
\text{high}_{u32}(x) := (x - \text{trunc}_{u32}(x)) \div_{\mathbb{Z}} 2^{32}.
\end{equation}

\begin{equation}
\text{high}_{s32} : \mathbb{Z} \to \mathbb{Z}.
\end{equation}
\begin{equation}
\text{high}_{s32}(x) := (x - \text{trunc}_{s32}(x)) \div_{\mathbb{Z}} 2^{32}.
\end{equation}

Let there be the following function, which computes the borrow of a 32-bit unsigned subtraction:
\begin{equation}
\text{borrow}_{u32} : 2^{32} \times 2^{32} \to 2^{32}.
\end{equation}
\begin{equation}
\text{borrow}_{u32}(x, y) = \begin{cases}
0 & \text{if}\ y \leq x, \\
1 & \text{if}\ y > x.
\end{cases}
\end{equation}

Let there be the following function, which computes the sign of an integer:
\begin{equation}
\text{sign} : \mathbb{Z} \to \mathbb{Z}.
\end{equation}
\begin{equation}
\text{sign}(x) := \begin{cases}
-1 & \text{if}\ x < 0, \\
0 & \text{if}\ x = 0, \\
1 & \text{if}\ x > 0.
\end{cases}
\end{equation}

Let there be the following function, which computes integer division, truncating towards zero:
\begin{equation}
\text{truncdiv} : \mathbb{Z} \times \mathbb{Z} \to \mathbb{Z}.
\end{equation}
\begin{equation}
\begin{array}{c}
\text{truncdiv}(x, y) := \text{the unique}\ z \in \mathbb{Z}\ \text{such that for some}\ r \in \mathbb{Z},\ 0 \leq |r| < |y|\ \text{and}\ x = y \times_{\mathbb{Z}} z +_{\mathbb{Z}} r, \\
\text{and}\ \text{sign}(r) = \text{sign}(x) \times_{\mathbb{Z}} \text{sign}(y).
\end{array}
\end{equation}

Let there be the following functions, which compute the truth values of integer equalities and inequalities:
\begin{equation}
=_{\mathbb{Z}}\ : \mathbb{Z} \times \mathbb{Z} \to \mathbb{Z}.
\end{equation}
\begin{equation}
=_{\mathbb{Z}}(x, y) := \begin{cases}
1 & \text{if}\ x = y, \\
0 & \text{otherwise}.
\end{cases}
\end{equation}

\begin{equation}
\leq_{\mathbb{Z}}\ : \mathbb{Z} \times \mathbb{Z} \to \mathbb{Z}
\end{equation}
\begin{equation}
\leq_{\mathbb{Z}}(x, y) := \begin{cases}
1 & \text{if}\ x \leq y, \\
0 & \text{otherwise}.
\end{cases}
\end{equation}

\begin{equation}
<_{\mathbb{Z}}\ : \mathbb{Z} \times \mathbb{Z} \to \mathbb{Z}
\end{equation}
\begin{equation}
<_{\mathbb{Z}}(x, y) := \begin{cases}
1 & \text{if}\ x < y, \\
0 & \text{otherwise}.
\end{cases}
\end{equation}

Let there be the following function, which decomposes a 32-bit integer into a bit vector:
\begin{equation}
\text{bits}_{u32} : 2^{32} \to (32 \to 2).
\end{equation}
Recall that 32 and 2 are von Neumann ordinals, being respectively the sets $\{0, ..., 31\}$ and $\{0,1\}$.
\begin{equation}
\text{bits}_{u32}(x) = \text{the unique}\ f : 32 \to 2\ \text{such that}\ x = \sum_{i=0}^{31} 2^i \times_{\mathbb{Z}} f(i).
\end{equation}

Let the following boolean operations be defined on bits:
\begin{equation}
\text{bitand} : 2 \times 2 \to 2
\end{equation}
\begin{equation}
\text{bitand}(x, y) = \begin{cases}
1 & \text{if}\ x = 1\ \text{and}\ y = 1, \\
0 & \text{otherwise}.
\end{cases}
\end{equation}

\begin{equation}
\text{bitor} : 2 \times 2 \to 2
\end{equation}
\begin{equation}
\text{bitor}(x, y) = \begin{cases}
1 & \text{if}\ x = 1\ \text{or}\ y = 1, \\
0 & \text{otherwise}.
\end{cases}
\end{equation}

\begin{equation}
\text{bitxor} : 2 \times 2 \to 2
\end{equation}
\begin{equation}
\text{bitor}(x, y) = \begin{cases}
1 & \text{if exactly one of}\ x\ \text{or}\ y\ \text{is 1}, \\
0 & \text{otherwise}.
\end{cases}
\end{equation}

Let the following function be defined:
\begin{equation}
\text{fmap}_{2\to 2^{32}} : (2 \times 2 \to 2) \to (2^{32} \times 2^{32} \to 2^{32}).
\end{equation}
\begin{equation}
\text{fmap}_{2\to 2^{32}}(f)(x, y) := \text{bits}_{u32}^{-1}(\lambda i : 32 \mapsto f(\text{bits}_{u32}(x)(i), \text{bits}_{u32}(y)(i))).
\end{equation}
In the above, $\lambda i : 32 \mapsto \cdots$ denotes the function which maps any $i \in 32$ to the value of $\cdots$ for that $i$.

\section{Transition function definition}

The transition function will execute the instruction pointed at by PC, and increment PC by one, unless one of the following is true:
\begin{enumerate}
\item The instruction at PC is a jump.
\item The machine is in a halted state.
\item PC is not aligned to an instruction boundary.
\item The instruction cannot be successfully executed given the current state.
\end{enumerate}
In conditions 2--4, the transition function maps the state to itself. This makes each of the conditions 3--4, where the machine is in a non-halted state, semantically equivalent to non-termination. Implementations may, in practice, halt execution with a useful error message when one of the conditions 3--4 occurs, instead of non-terminating.

The transition function $\tau$ has type
\begin{equation}
\tau : \Sigma \to \Sigma.
\end{equation}
Recall that $\Sigma$ is defined as a Cartesian product:
\begin{equation}
\Sigma := \prod_{i \in \text{Chips}} \Sigma_i.
\end{equation}
$\tau$ can be defined as a Cartesian product of functions:
\begin{equation}
\tau := \prod_{i \in \text{Chips}} \tau_i.
\end{equation}
For all $i \in \text{Chips}$:
\begin{equation}
\tau_i : \Sigma \to \Sigma_i.
\end{equation}
For all $i \in \text{Chips} \setminus \{\text{CPU, Program, StaticData, Memory, Output}\}$, $\Sigma_i \cong 1$ and therefore $\tau_i$ can be defined as the unique function $\tau_i : \Sigma \to \Sigma_i$. So to define $\tau$, it suffices to define the following functions:
\begin{equation}
\tau_{\text{CPU}} : \Sigma \to \Sigma_{\text{CPU}},
\end{equation}
\begin{equation}
\label{eq:tau-program-type}
\tau_{\text{Program}} : \Sigma \to \Sigma_{\text{Program}},
\end{equation} 
\begin{equation}
\label{eq:tau-static-data-type}
\tau_{\text{StaticData}} : \Sigma \to \Sigma_{\text{StaticData}},
\end{equation}
\begin{equation}
\tau_{\text{Memory}} : \Sigma \to \Sigma_{\text{Memory}},
\end{equation}
\begin{equation}
\tau_{\text{Output}} : \Sigma \to \Sigma_{\text{Output}}.
\end{equation}

Of these, $\tau_{\text{Program}}$ and $\tau_{\text{StaticData}}$ are easiest to define, because these chips' states are immutable:
\begin{equation}
\tau_{\text{Program}} := \pi_{\text{Program}}.
\end{equation}
\begin{equation}
\tau_{\text{StaticData}} := \pi_{\text{StaticData}}.
\end{equation}

For the remaining chips, namely CPU, Memory, and Output, the result of the transition function depends on which instruction is located at the current PC. These functions are defined as follows, for all $i \in \{\text{CPU, Memory, Output}\}$:

\begin{equation}
\tau_i(s) = \begin{cases}
s & \text{fetch}(s) \in \text{FetchError}, \\
\tau_{i,\text{fetch}(s)}(s) & \text{fetch}(s) \in \text{Instruction}.
\end{cases}
\end{equation}

The fetch function maps a state to the instruction at the current PC, or an error. Its type is as follows:
\begin{equation}
\text{fetch} : \Sigma \to (\text{Instruction} \cup \text{FetchError}).
\end{equation}
\begin{equation}
\text{FetchError} := \{\text{PCNotDefined}\}.
\end{equation}
The set FetchError is assumed to be disjoint from Instruction. The error PCNotDefined indicates that there is no instruction in the current program, at the current value of PC.
\begin{equation}
\text{fetch}(s) := \begin{cases}
\pi_{\text{Program}}(s)(pc \div_{\mathbb{Z}} 24) & \text{if}\ \pi_{\text{Program}}(s)(pc \div_{\mathbb{Z}} 24)\ \text{is defined}, \\
\text{PCNotDefined} & \text{otherwise},
\end{cases}
\end{equation}
where $pc := \pi_{\text{PC}}(\pi_{\text{CPU}}(s))$.

The functions $\tau_{i,j}$, for all $i \in \{\text{CPU, Memory, Output}\}$ and all $j \in \text{Instruction}$, have the following types:
\begin{equation}
\tau_{i,j} : \Sigma \to \Sigma_i.
\end{equation}

\begin{figure}
\begin{tabular}{llll}
\textbf{Opcode} ($\pi_{1}(i)$) & \textbf{Left} & \textbf{Right} & \textbf{Operator} ($\phi_i$) \\
Add & Var & Var & $\phi_i(x,y) := \pi_{u32}^{-1}(\text{trunc}_{u32}(\pi_{u32}(x) +_{\mathbb{Z}} \pi_{u32}(x)))$ \\
Addi & Var & Imm & $\phi_i(x,y) := \pi_{u32}^{-1}(\text{trunc}_{u32}(\pi_{u32}(x) +_{\mathbb{Z}} \pi_{u32}(x)))$ \\
Addc & Var & Var & $\phi_i(x,y) := \pi_{u32}^{-1}(\text{high}_{u32}(\pi_{u32}(x) +_{\mathbb{Z}} \pi_{u32}(x)))$ \\
Addci & Var & Imm & $\phi_i(x,y) := \pi_{u32}^{-1}(\text{high}_{u32}(\pi_{u32}(x) +_{\mathbb{Z}} \pi_{u32}(x)))$ \\
Sub & Var & Var & $\phi_i(x,y) := \pi_{u32}^{-1}(\text{trunc}_{u32}(\pi_{u32}(x) -_{\mathbb{Z}} \pi_{u32}(x)))$ \\
Subi & Var & Imm & $\phi_i(x,y) := \pi_{u32}^{-1}(\text{trunc}_{u32}(\pi_{u32}(x) -_{\mathbb{Z}} \pi_{u32}(x)))$ \\
Isub & Imm & Var & $\phi_i(x,y) := \pi_{u32}^{-1}(\text{trunc}_{u32}(\pi_{u32}(x) -_{\mathbb{Z}} \pi_{u32}(x)))$ \\
Subb & Var & Var & $\phi_i(x,y) := \text{borrow}_{u32}(x, y)$ \\
Subbi & Var & Imm & $\phi_i(x,y) := \text{borrow}_{u32}(x, y)$ \\
Isubb & Imm & Var & $\phi_i(x,y) := \text{borrow}_{u32}(x, y)$ \\
Mul & Var & Var & $\phi_i(x,y) := \pi_{u32}^{-1}(\text{trunc}_{u32}(\pi_{u32}(x) \times_{\mathbb{Z}} \pi_{u32}(y)))$ \\
Muli & Var & Imm & $\phi_i(x,y) := \pi_{u32}^{-1}(\text{trunc}_{u32}(\pi_{u32}(x) \times_{\mathbb{Z}} \pi_{u32}(y)))$ \\
Mulhs & Var & Var & $\phi_i(x,y) := \pi_{s32}^{-1}(\text{high}_{s32}(\pi_{s32}(x) \times_{\mathbb{Z}} \pi_{s32}(y)))$ \\
Mulhsi & Var & Imm & $\phi_i(x,y) := \pi_{s32}^{-1}(\text{high}_{s32}(\pi_{s32}(x) \times_{\mathbb{Z}} \pi_{s32}(y)))$ \\
Mulhu & Var & Var & $\phi_i(x,y) := \pi_{u32}^{-1}(\text{high}_{u32}(\pi_{u32}(x) \times_{\mathbb{Z}} \pi_{u32}(y)))$ \\
Mulhui & Var & Imm & $\phi_i(x,y) := \pi_{u32}^{-1}(\text{high}_{u32}(\pi_{u32}(x) \times_{\mathbb{Z}} \pi_{u32}(y)))$ \\
Div & Var & Var & $\phi_i(x,y) := \pi_{u32}^{-1}(\text{truncdiv}(\pi_{u32}(x), \pi_{u32}(y)))$ \\
Divi & Var & Imm & $\phi_i(x,y) := \pi_{u32}^{-1}(\text{truncdiv}(\pi_{u32}(x), \pi_{u32}(y)))$ \\
Sdiv & Var & Var & $\phi_i(x,y) := \pi_{s32}^{-1}(\text{truncdiv}(\pi_{s32}(x), \pi_{s32}(y)))$ \\
Sdivi & Var & Imm & $\phi_i(x,y) := \pi_{s32}^{-1}(\text{truncdiv}(\pi_{s32}(x), \pi_{s32}(y)))$ \\
Shl & Var & Var & $\phi_i(x,y) := \pi_{u32}^{-1}(\text{trunc}_{u32}(\pi_{u32}(x) \times_{\mathbb{Z}} 2^{\text{trunc}_{u5}(\pi_{u32}(y))}))$ \\
Shli & Var & Imm & $\phi_i(x,y) := \pi_{u32}^{-1}(\text{trunc}_{u32}(\pi_{u32}(x) \times_{\mathbb{Z}} 2^{\text{trunc}_{u5}(\pi_{u32}(y))}))$ \\
Ishl & Imm & Var &  $\phi_i(x,y) := \pi_{u32}^{-1}(\text{trunc}_{u32}(\pi_{u32}(x) \times_{\mathbb{Z}} 2^{\text{trunc}_{u5}(\pi_{u32}(y))}))$ \\
Shr & Var & Var & $\phi_i(x,y) := \pi_{u32}^{-1}(\text{truncdiv}(\pi_{u32}(x), 2^{\text{trunc}_{u5}(\pi_{u32}(y))}))$ \\
Shri & Var & Imm & $\phi_i(x,y) := \pi_{u32}^{-1}(\text{truncdiv}(\pi_{u32}(x), 2^{\text{trunc}_{u5}(\pi_{u32}(y))}))$ \\
Ishr & Imm & Var &  $\phi_i(x,y) := \pi_{u32}^{-1}(\text{truncdiv}(\pi_{u32}(x), 2^{\text{trunc}_{u5}(\pi_{u32}(y))}))$ \\
Sra & Var & Var & $\phi_i(x,y) := \pi_{s32}^{-1}(\text{truncdiv}(\pi_{s32}(x), 2^{\text{trunc}_{u5}(\pi_{u32}(y))}))$ \\
Srai & Var & Imm & $\phi_i(x,y) := \pi_{s32}^{-1}(\text{truncdiv}(\pi_{s32}(x), 2^{\text{trunc}_{u5}(\pi_{u32}(y))}))$ \\
Isra & Imm & Var & $\phi_i(x,y) := \pi_{s32}^{-1}(\text{truncdiv}(\pi_{s32}(x), 2^{\text{trunc}_{u5}(\pi_{u32}(y))}))$ \\
\end{tabular}
\caption{Binary function opcodes (1 of 2)}
\label{tbl:binary-function-opcodes-1}
\end{figure}

\begin{figure}
\begin{tabular}{llll}
Lt & Var & Var & $\phi_i(x,y) := \pi_{u32}^{-1}(<_{\mathbb{Z}}(\pi_{u32}(x), \pi_{u32}(y)))$ \\
Lti & Var & Imm & $\phi_i(x,y) := \pi_{u32}^{-1}(<_{\mathbb{Z}}(\pi_{u32}(x), \pi_{u32}(y)))$ \\
Ilt & Imm & Var & $\phi_i(x,y) := \pi_{u32}^{-1}(<_{\mathbb{Z}}(\pi_{u32}(x), \pi_{u32}(y)))$ \\
Lte & Var & Var & $\phi_i(x,y) := \pi_{u32}^{-1}(\leq_{\mathbb{Z}}(\pi_{u32}(x), \pi_{u32}(y)))$ \\
Ltei & Var & Imm & $\phi_i(x,y) := \pi_{u32}^{-1}(\leq_{\mathbb{Z}}(\pi_{u32}(x), \pi_{u32}(y)))$ \\
Ilte & Imm & Var &  $\phi_i(x,y) := \pi_{u32}^{-1}(\leq_{\mathbb{Z}}(\pi_{u32}(x), \pi_{u32}(y)))$ \\
Slt & Var & Var & $\phi_i(x,y) := \pi_{u32}^{-1}(<_{\mathbb{Z}}(\pi_{s32}(x), \pi_{s32}(y)))$ \\
Slti & Var & Imm & $\phi_i(x,y) := \pi_{u32}^{-1}(<_{\mathbb{Z}}(\pi_{s32}(x), \pi_{s32}(y)))$ \\
Islt & Imm & Var & $\phi_i(x,y) := \pi_{u32}^{-1}(<_{\mathbb{Z}}(\pi_{s32}(x), \pi_{s32}(y)))$ \\
Slte & Var & Var & $\phi_i(x,y) := \pi_{u32}^{-1}(\leq_{\mathbb{Z}}(\pi_{s32}(x), \pi_{s32}(y)))$ \\
Islte & Imm & Var & $\phi_i(x,y) := \pi_{u32}^{-1}(\leq_{\mathbb{Z}}(\pi_{s32}(x), \pi_{s32}(y)))$ \\
Sltei & Var & Imm & $\phi_i(x,y) := \pi_{u32}^{-1}(\leq_{\mathbb{Z}}(\pi_{s32}(x), \pi_{s32}(y)))$ \\
Eq & Var & Var & $\phi_i(x,y) := \pi_{u32}^{-1}(=_{\mathbb{Z}}(\pi_{u32}(x), \pi_{u32}(y)))$ \\
Eqi & Var & Imm & $\phi_i(x,y) := \pi_{u32}^{-1}(=_{\mathbb{Z}}(\pi_{u32}(x), \pi_{u32}(y)))$ \\
Ne & Var & Var & $\phi_i(x,y) := \pi_{u32}^{-1}(1 - =_{\mathbb{Z}}(\pi_{u32}(x), \pi_{u32}(y)))$ \\
Nei & Var & Imm & $\phi_i(x,y) := \pi_{u32}^{-1}(1 - =_{\mathbb{Z}}(\pi_{u32}(x), \pi_{u32}(y)))$ \\
And & Var & Var & $\phi_i := \text{fmap}_{2\to 2^{32}}(\text{bitand})$ \\
Andi & Var & Imm & $\phi_i := \text{fmap}_{2\to 2^{32}}(\text{bitand})$ \\
Or & Var & Var & $\phi_i := \text{fmap}_{2\to 2^{32}}(\text{bitor})$ \\
Ori & Var & Imm & $\phi_i := \text{fmap}_{2\to 2^{32}}(\text{bitor})$ \\
Xor & Var & Var & $\phi_i := \text{fmap}_{2\to 2^{32}}(\text{bitxor})$ \\
Xori & Var & Imm & $\phi_i := \text{fmap}_{2\to 2^{32}}(\text{bitxor})$ 
\end{tabular}
\caption{Binary function opcodes (2 of 2)}
\label{tbl:binary-function-opcodes-2}
\end{figure}

Most Valida opcodes perform a binary operation on two inputs, 1--2 of which are stack variables, and 0--1 of which are immediate values, with a single output which is stored in a stack variable. Figures~\ref{tbl:binary-function-opcodes-1}~and~\ref{tbl:binary-function-opcodes-2} enumerate these binary function opcodes, specifying which inputs are variable and which are immediate, and which binary function is used to compute the output from the input. The transition functions for instructions $i \in \text{Instruction}$ with binary function opcodes are defined generically as follows.

\begin{equation}
\label{eq:cpu-transition-non-jump}
\tau_{\text{CPU},i}(s) := (\pi_{\text{PC}}(s') +_{\mathbb{Z}_{2^{32}}} 1, \pi_{\text{FP}}(s'), \pi_{\text{UnconsumedInput}}(s'), \pi_{\text{Halting}}(s')),\ \text{where}\ s' = \pi_{\text{CPU}}(s).
\end{equation}

The above indicates that a binary function instruction acts on the CPU state by simply incrementing PC (with wrap-around) and doing nothing else. It is a fact (provable by induction on the length of an execution) that if the program is in a halted state, then the current instruction is a STOP opcode, and hence not a binary function opcode. This means that the final term in the above definition, $\pi_{\text{Halting}}(s)$, could be replaced with $\text{NotHalted}$ without changing the semantics.

\begin{equation}
\label{eq:output-projection}
\tau_{\text{Output},i}(s) := \pi_{\text{Output}}(s).
\end{equation}

The above indicates that a binary function instruction acts on the output tape state by doing nothing to it. The definition of how a binary function instruction acts on memory is more complex. It depends on the binary function's operand types (stack variable or immediate), as well as the binary function for the instruction opcode.

The following function can be used to get the value denoted by an input operand, based on its type as determined by the opcode, its value, and the state:
\begin{equation}
\iota : \{\text{Var, Imm}\} \times 2^{32} \times \Sigma \to 2^{32}.
\end{equation}
\begin{equation}
\iota(\text{Imm}, v, s) := v.
\end{equation}
\begin{equation}
\iota(\text{Var}, v, s) := \text{load}(s, \pi_{\text{FP}}(\pi_{\text{CPU}}(s)) +_{\mathbb{Z}_{2^{32}}} v).
\end{equation}

The following function can be used to denote a memory state where the stack operand with the specified FP offset is updated to the specified value:
\begin{equation}
\text{Update} : \Sigma \to 4(2^{30}) \to 2^{32} \to \Sigma_{\text{Memory}}.
\end{equation}
\begin{equation}
\text{Update}(s, v, x) := \text{store}(s, \pi_{\text{FP}}(\pi_{\text{CPU}}(s)) + v, x).
\end{equation}

Let $i$ be an instruction with a binary function opcode. Let $\phi_i : 2^{32} \times 2^{32} \to 2^{32}$ be the binary function corresponding to the opcode $\pi_{\text{OP}}(i)$ of instruction $i$, as enumerated in Figure~\ref{tbl:binary-function-opcodes-1} and Figure~\ref{tbl:binary-function-opcodes-2}. Let $t_{i,1} \in \{\text{Var, Imm}\}$ be the first operand type of instruction $i$, as indicated in the same figures. Let $t_{i,2}$ be the second operand type of instruction $i$, as indicated in the same figures. Then the memory transition function $\tau_{\text{Memory},i}$ is defined as follows:

\begin{equation}
\tau_{\text{Memory},i}(s) := \text{Update}(s, \pi_1(i), \phi_i(\iota(t_{i,1}, \pi_2(i), s), \iota(t_{i,2}, \pi_3(i), s))).
\end{equation}

That completes the definition of the transition function for binary function opcodes. The remaining (i.e., non binary function) opcodes are all and only the CPU and Output chip opcodes, namely: Store32, StoreU8, Load32, LoadU8, LoadS8, Jal, Jalv, Beq, Beqi, Bne, Bnei, Imm32, ReadAdvice, Stop, LoadFp, and Write. To define the transition function for these remaining opcodes, it suffices to define the CPU, Memory, and Output chip transition functions for these opcodes.

The non binary function opcodes can be further subcategorized into the jump and non-jump opcodes. The jump opcodes are Jal, Jalv, Beq, Beqi, Bne, and Bnei. The non-jump opcodes are Store32, StoreU8, Load32, LoadU8, LoadS8, Imm32, ReadAdvice, LoadFp, Stop, and Write. The jump opcodes can modify FP, and they can modify PC (other than by incrementing it), but they do not modify memory. The non-jump opcodes can modify memory, but they cannot modify FP, and they cannot modify PC (other than by incrementing it).

For all instructions $i$ with opcodes of StoreU8, LoadU8, LoadS8, Imm32, ReadAdvice, LoadFp, and Write, the CPU chip transition function is defined in Equation~\ref{eq:cpu-transition-non-jump}.

For all instructions $i$ with opcode Store32, the CPU chip transition function is defined as:
\begin{equation}
\tau_{\text{CPU},i}(s) := \begin{cases}
(\pi_{\text{PC}}(s') +_{\mathbb{Z}_{2^{32}}} 1, \pi_{\text{FP}}(s'), \pi_{\text{UnconsumedInput}}(s'), \pi_{\text{Halting}}(s')) & \text{if}\ a \in 4(2^{30}), \\
s' & \text{otherwise},
\end{cases}
\end{equation}
where $s' := \pi_{\text{CPU}}(s)$ and $a = \iota(\text{Var}, \pi_2(i), s)$.

For all instructions $i$ with opcode Load32, the CPU chip transition function is defined as:
\begin{equation}
\tau_{\text{CPU},i}(s) := \begin{cases}
(\pi_{\text{PC}}(s') +_{\mathbb{Z}_{2^{32}}} 1, \pi_{\text{FP}}(s'), \pi_{\text{UnconsumedInput}}(s'), \pi_{\text{Halting}}(s')) & \text{if}\ a \in 4(2^{30}), \\
s' & \text{otherwise},
\end{cases}
\end{equation}
where $s' = \pi_{\text{CPU}}(s)$ and $a = \iota(\text{Var}, \pi_3(i), s)$.

For all instructions with opcodes of Beq, Beqi, Bne, or Bnei, the CPU chip transition function is defined as:
\begin{equation}
\tau_{\text{CPU},i}(s) := (p, \pi_{\text{FP}}(s'), \pi_{\text{UnconsumedInput}}(s'), \pi_{\text{Halting}}(s')),
\end{equation}
where
\begin{equation}
s' = \pi_{\text{CPU}}(s),
\end{equation}
\begin{equation}
p = \begin{cases}
\pi_{\text{PC}}(s') +_{\mathbb{Z}_{2^{32}}} 1 & \text{if}\ c = 0, \\
\pi_{u32}^{-1}(\text{truncdiv}(\pi_{u32}(\pi_1(i)), 24)) & \text{if}\ c = 1,
\end{cases}
\end{equation}
\begin{equation}
c = \begin{cases}
x =_{\mathbb{Z}} y & \text{if Opcode}_i \in \{\text{Beq, Beqi}\}, \\
1 - (x =_{\mathbb{Z}} y) & \text{if Opcode}_i \in \{\text{Bne, Bnei}\},
\end{cases}
\end{equation}
\begin{equation}
x = \iota(\text{Var}, \pi_2(i), s),
\end{equation}
\begin{equation}
y = \iota(t_y, \pi_3(i), s),
\end{equation}
\begin{equation}
t_y = \begin{cases}
\text{Var} & \text{if Opcode}_i \in \{\text{Beq, Bne}\}, \\
\text{Imm} & \text{if Opcode}_i \in \{\text{Beqi, Bnei}\}.
\end{cases}
\end{equation}

For all instructions $i$ with opcode Stop, the CPU chip transition function is defined as:
\begin{equation}
\tau_{\text{CPU},i}(s) := (\pi_{\text{PC}}(s'), \pi_{\text{FP}}(s'), \pi_{\text{UnconsumedInput}}(s'), \text{Halted}),\ \text{where}\ s' = \pi_{\text{CPU}}(s).
\end{equation}

For all instructions $i$ with opcode Jal, the CPU chip transition function is defined as:
\begin{equation}
\tau_{\text{Jal},i}(s) := (a, \pi_{\text{FP}}(s') +_{\mathbb{Z}_{2^{32}}} \pi_3(i), \pi_{\text{UnconsumedInput}}(s'), \pi_{\text{Halting}}(s')),
\end{equation}
where
\begin{equation}
a := \pi_{u32}^{-1}(\text{truncdiv}(\pi_{u32}(\pi_2(i)), 24)),
\end{equation}
\begin{equation}
s' := \pi_{\text{CPU}}(s).
\end{equation}

For all instructions $i$ with opcode Jalv, the CPU chip transition function is defined as:
\begin{equation}
\tau_{\text{Jalv},i}(s) := (\iota(\text{Var}, \pi_2(i), s), \iota(\text{Var}, \pi_1(i), s), \pi_{\text{UnconsumedInput}}(s'), \pi_{\text{Halting}}(s')),
\end{equation}
where
\begin{equation}
s' = \pi_{\text{CPU}}(s).
\end{equation}

For all instructions $i$ with opcode ReadAdvice, the CPU chip transition function is defined as:
\begin{equation}
\tau_{\text{CPU},i}(s) := (\pi_{\text{PC}}(s') +_{\mathbb{Z}^{2^{32}}} 1, \pi_{\text{FP}}(s'), \text{tail}'(\pi_{\text{UnconsumedInput}}(s')), \pi_{\text{Halting}}(s')),\ \text{where}\ s' = \pi_{\text{CPU}}(s).
\end{equation}

For all instructions $i$ with opcode ReadAdvice, the memory chip transition function is defined as:
\begin{equation}
\tau_{\text{Memory},i}(s) := \text{Update}(s, \pi_1(i), c),\ \text{where}
\end{equation}
\begin{equation}
c := \begin{cases}
\text{head}(\pi_{\text{UnconsumedInput}}(\pi_{\text{CPU}}(s))), & \text{if}\ \pi_{\text{UnconsumedInput}}(\pi_{\text{CPU}}(s))\ \text{is non-empty}, \\
2^{32}-1 & \text{otherwise}.
\end{cases}
\end{equation}

For all instructions $i$ with opcodes of Beq, Beqi, Bne, Bnei, Stop, or Write, the Memory chip transition function is just the projection function, because the opcode does not modify memory:
\begin{equation}
\label{eq:memory-projections}
\tau_{\text{Memory},i} := \pi_{\text{Memory}}.
\end{equation}

For all instructions $i$ with opcode of LoadFp, the Memory chip transition function is:
\begin{equation}
\tau_{\text{Memory},i}(s) := \text{Update}(s, \pi_1(i), \pi_{\text{FP}}(\pi_{\text{CPU}}(s)) + \pi_2(i)).
\end{equation}

For all instructions $i$ with opcode of Imm32, the Memory chip transition function is:
\begin{equation}
\tau_{\text{Memory},i}(s) := \text{Update}(s, \pi_1(i), \sum_{j=0}^3 2^j \times_{\mathbb{Z}_{2^{32}}} \pi_{2+j}(i)).
\end{equation}

For all instructions $i$ with opcode of Store32, the Memory chip transition function is:
\begin{equation}
\tau_{\text{Memory},i}(s) := \begin{cases}
\text{store}(s, \text{load}(s, a), \iota(\text{Var}, \pi_3(i), s)) & \text{if}\ a \in 4(2^{30}), \\
\pi_{\text{Memory}}(s) & \text{otherwise}.
\end{cases}
\end{equation}
where
\begin{equation}
a := \iota(\text{Var}, \pi_2(i), s).
\end{equation}

For all instructions $i$ with opcode of StoreU8, the Memory chip transition function is:
\begin{equation}
\tau_{\text{Memory},i}(s) := \pi_{\text{Memory}}(s)[\iota(\text{Var}, \pi_2(i), s) \mapsto \pi_{\text{Memory}}(s)(\iota(\text{Var}, \pi_3(i), s))].
\end{equation}

For all instructions $i$ with opcode of Load32, the Memory chip transition function is:
\begin{equation}
\tau_{\text{Memory},i}(s) := \begin{cases}
\text{Update}(s, \pi_1(i), \text{load}(s, a)) & \text{if}\ a \in 4(2^{30}), \\
\pi_{\text{Memory}}(s) & \text{otherwise},
\end{cases}
\end{equation}
where
\begin{equation}
a := \iota(\text{Var}, \pi_2(i), s).
\end{equation}

For all instructions $i$ with opcode of LoadU8, the Memory chip transition function is:
\begin{equation}
\tau_{\text{Memory},i}(s) := \text{Update}(s, \pi_1(i), \pi_{u32}^{-1}(\pi_{u8}(\pi_{\text{Memory}}(s)(\iota(\text{Var}, \pi_3(i), s))))).
\end{equation}

For all instructions $i$ with opcode of LoadS8, the Memory chip transition function is:
\begin{equation}
\tau_{\text{Memory},i}(s) := \text{Update}(s, \pi_1(i), \pi_{s32}^{-1}(\pi_{s8}(\pi_{\text{Memory}}(s)(\iota(\text{Var}, \pi_3(i), s))))).
\end{equation}

For all instructions $i$ with opcode of Jal or Jalv, the Memory chip transition function is:
\begin{equation}
\tau_{\text{Memory},i}(s) := \text{store}(s, \iota(\text{Var}, \pi_1(i), s), \pi_{u32}^{-1}(24 \times_{\mathbb{Z}} (\pi_{u32}(\pi_{\text{PC}}(\pi_{\text{CPU}}(s))) +_{\mathbb{Z}} 1))).
\end{equation}

For all instructions $i$ with opcodes other than Write, the Output chip transition function is just the projection function, because the opcode does not modify the output. See Equation~\ref{eq:output-projection} for the definition of the Output chip transition function for instructions $i$ with opcodes other than Write.

For instructions $i$ with opcode Write, the Output chip transition function is:
\begin{equation}
\tau_{\text{Output},i}(s) :=
\pi_{\text{Output}(s)} +^* (\text{trunc}_{u8}(\iota(\text{Var}, \pi_1(i), s))).
\end{equation}

\bibliographystyle{plain}
\bibliography{refs}

\begin{thebibliography}{10}

\bibitem{zexe}
Sean Bowe, Alessandro Chiesa, Matthew Green, Ian Miers, Pratyush Mishra, and
  Howard Wu.
\newblock {Zexe: Enabling Decentralized Private Computation}.
\newblock Cryptology ePrint Archive, Paper 2018/962, 2018.
\newblock \url{https://eprint.iacr.org/2018/962.pdf}.

\bibitem{leo}
Collin Chin, Howard Wu, Raymond Chu, Alessandro Coglio, Eric McCarthy, and Eric
  Smith.
\newblock {Leo: A Programming Language for Formally Verified, Zero-Knowledge
  Applications}.
\newblock Cryptology ePrint Archive, Paper 2021/651, 2021.
\newblock \url{https://eprint.iacr.org/2021/651.pdf}.

\bibitem{zksync}
The~ZKsync Community.
\newblock {ZKsync Protocol}.
\newblock \url{https://docs.zksync.io/zksync-protocol}.

\bibitem{validaspec}
Max Gillett, Daniel Lubarov, and Wei Dai.
\newblock {Working ISA Spec}.
\newblock \url{https://github.com/valida-xyz/valida-compiler/issues/2}, 2023.

\bibitem{opsuccinct}
Edward Li.
\newblock {Introducing OP Succinct Lite: ZK Fraud Proofs on the OP Stack}.
\newblock \url{https://blog.succinct.xyz/op-succinct-lite/?ref=conduit.xyz},
  February 2025.

\bibitem{litabench}
Lita.
\newblock {Benchmarks}.
\newblock
  \url{https://lita.gitbook.io/lita-documentation/architecture/benchmarks},
  August 2024.

\bibitem{validadocs}
Lita.
\newblock {Lita Docs}.
\newblock \url{https://lita.gitbook.io/lita-documentation}, 2025.

\bibitem{starkex}
StarkEx.
\newblock {StarkEx Docs}.
\newblock \url{https://docs.starkware.co/starkex/index.html}, 2023.

\bibitem{cairo}
Starkware.
\newblock Cairo.
\newblock \url{https://starkware.co/cairo/}, 2025.

\bibitem{thomas25}
Morgan Thomas.
\newblock {Optimizing the Ethereum Execution Engine for Succinct Proofs With
  Valida}.
\newblock
  \texttt{https://www.lita.foundation/blog/optimizing-the-ethereum-execution-engine-for-\allowbreak
  succinct-proofs-with-valida}, April 2025.

\bibitem{yu24}
Whisker Yu.
\newblock {How to Develop ZK Fraud Proof with RISC0}.
\newblock
  \url{https://0xwhisker.hashnode.dev/how-to-develop-zk-fraud-proof-with-risc0},
  January 2024.

\end{thebibliography}

\end{document}